\documentclass{iau_FM}
\usepackage{graphicx}

\def\apj{ApJ}

\def\an{AN}
\def\mnras{MNRAS}

\def\apjl{ApJL}

\def\nat{Nature}

\def\araa{ARAA}

\newcommand{\Lar}{r_{\rm L}}
\renewcommand{\vec}[1]{\mathbf{#1}}	%vector
\newcommand{\dd}{\mathrm{d}}        %differential
\newcommand{\pc}{\,{\rm pc}}     %parsec
\newcommand{\kpc}{\,{\rm kpc}}  %kpc
\newcommand{\brms}{\,b_{\rm rms}}

\newcommand{\ncr}{n_{\rm cr}}
\newcommand{\ecr}{e_{\rm cr}}
\newcommand{\ncre}{n_{\rm cr}^{\rm(e)}}

\title[On energy equipartition between cosmic rays and magnetic fields] %% give here short title %%
{On energy equipartition between cosmic rays and magnetic fields}

\author[A. Seta et al.]   %% give here short author list %%
{Amit Seta, Anvar Shukurov, Paul J.\ Bushby \& Toby S.\ Wood}

\affiliation{School of Mathematics, Statistics \& Physics, Newcastle University, UK \\email: {\tt a.seta1@ncl.ac.uk}}

\pubyear{2018}
\jname{Astronomy in Focus, Volume 1} 
\editors{L. Feretti, G. Heald, M. Johnston-Hollitt \& F. Govoni, ed.}
\begin{document}
\maketitle
\begin{abstract}
Interpretations of synchrotron observations
often assume a tight correlation between magnetic and cosmic ray energy densities.
We examine this
assumption 
using both test-particle simulations of cosmic rays
and MHD simulations which include cosmic rays as a
diffusive fluid.
We find no spatial correlation between the cosmic rays and magnetic field energy densities at turbulent scales. 
Moreover, the cosmic ray number density and magnetic field energy density are statistically independent. Nevertheless, 
the cosmic ray spatial distribution is
highly inhomogeneous, especially at low energies because
the particles are trapped between random magnetic mirrors.
These results can significantly change the interpretation of synchrotron observations and thus our understanding of the strength and structure of magnetic fields in the Milky Way and nearby spiral galaxies.
\keywords{Cosmic rays, Magnetic fields, Diffusion, Radio continuum: ISM}
\end{abstract}
%--------------------------------------------------------------------------------------
\firstsection % if your document starts with a section,  % remove some space above using this command.
\section{Introduction}  \label{intro}
Synchrotron emission
is one of the main observational probes of galactic magnetic fields.
The synchrotron intensity $I$  depends on
the number density of 
cosmic ray electrons $\ncre$ and %
magnetic field $b_{\perp}$
perpendicular to the line of sight $\vec{\ell}$,
\begin{equation} \label{syn}
I = K \int_L \ncre \, b_{\perp}^{(p+1)/2} \, \dd\ell\,,
\end{equation}
where $K$ is a constant, 
%(which depends on emission frequency), 
$L$ is the total path length 
and $p$ is the power-law index of the electron energy spectrum ($p\approx3$ in spiral galaxies).
The determination of magnetic field strength from the synchrotron intensity requires an independent observation of
the number density of cosmic ray electrons. In the absence of such information, an energy equipartition between the total cosmic ray 
(consisting of $90 \%$ protons, 1--2\% electrons and the rest are heavier particles) and magnetic field energy densities is assumed in order to 
extract the magnetic field strength.  Though there is no convincing theoretical justification or observational evidence to assume
point-to-point energy equipartition between cosmic rays and magnetic fields in spiral galaxies, it
is widely used to obtain magnetic field strengths from synchrotron intensities (\cite{Beck2005}).
%-----------------------------------------------------
\section{Energy equipartition argument: previous tests and our approach} \label{prev}
The energy equipartition assumption was first used to study the energy content in cosmic rays and magnetic fields in the jet of M87
using optical and radio emissions from the system (\cite{B56}). Since then, the energy equipartition assumption has been used to obtain the magnetic field strength in various systems. 
\cite{Duric90} suggested that the range of possible
magnetic field strengths in spiral galaxies can differ by at most 
an order of magnitude from the equipartition value.
The conclusion is justified as follows. If the magnetic fields are considerably weaker than the equipartition value then the particles would quickly escape the disk and
almost no synchrotron emission would be observed. On the other hand, if the fields are far stronger, then the particles would be confined very close to the sources and strong
emission would be observed but only in small regions around the sources. Neither of the above situations is actually seen in spiral galaxies.
Also, a weakness of this type of argument is that cosmic ray diffusivity depends not on the magnetic field
strength but on the \textit{ratio} of the random to large-scale magnetic field strengths.
\cite{CW93} used gamma-ray observations 
to obtain the proton number density in the Large and 
Small Magellanic Clouds (LMC and SMC). They showed that energy equipartition does not hold for these irregular galaxies.
However, \cite{Mao2012}, also using gamma-ray data, 
concluded that the equipartition seems to hold
in the LMC. 
The radio-FIR correlation studies suggest
that the energy equipartition assumption is invalid on scales
smaller than a few $\kpc$ and may hold on larger scales (\cite{BR13}). \cite{YGZ16} analyzed
the gamma-ray and radio spectra 
of starburst galaxies and concluded that equipartition does not hold in these dynamic systems.
The energy equipartition assumption, when applied on large scales, seems to hold for a number of systems (\cite{Beck2016}). However, there are also
cases where this is not true.  Here, using test--particle and magnetohydrodynamic (MHD) simulations, we test the point-wise or local energy equipartition assumption.
%-------------------------------------------------------
\section{Cosmic rays as test particles} \label{crpart}
\begin{figure}%[h]
        \includegraphics[width=\columnwidth]{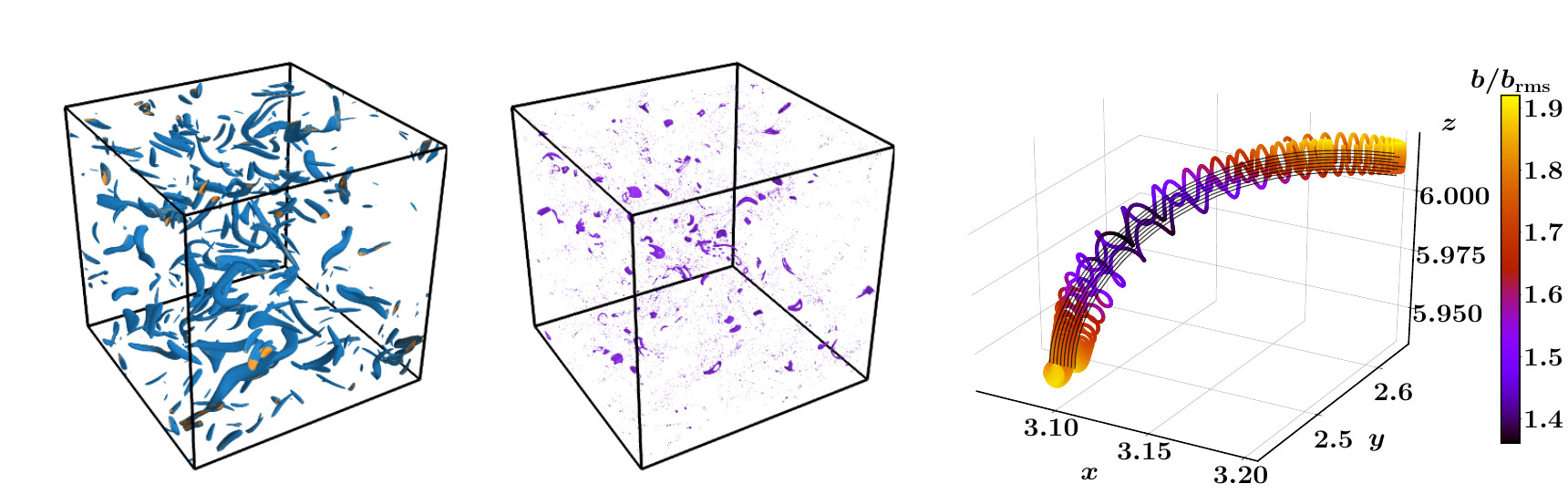} 
	\caption{Left: The normalised strength of the intermittent magnetic field
$b^2/\langle b^2 \rangle =12$ (blue), $15$ (yellow) from a fluctuation dynamo simulation ($\langle\cdots \rangle$ denotes a volume average).
Middle: The normalised cosmic ray number density $\ncr/\langle \ncr \rangle = 3.5$ for low-energy particles
($\Lar/l_0 = 0.0016$).
The magnetic field and cosmic ray number density are uncorrelated and yet there are small-scale structures in the cosmic ray distribution.
This is due to magnetic traps which depends on the topology of magnetic field lines rather than the field strength. Right: The trajectory of a trapped particle in the simulation, with the magnetic field strength along the trajectory shown with colour. 
The dark grey lines show magnetic field lines. The particle moves forward and backward between two magnetic mirrors.}
        \label{fig_particles}
\end{figure}

Cosmic ray particles gyrate around the magnetic fields lines and are scattered by small-scale magnetic fluctuations, which leads to their diffusion in galaxies (\cite{Cesarsky1980}).
Such fluctuations can be spatially Gaussian or intermittent. We generate an intermittent magnetic field by numerically solving the induction
equation using a chaotic velocity field with the Kolmogorov energy spectrum. The induction equation is solved using a finite difference numerical scheme 
in a periodic box of dimensionless size $2 \pi$ (physically equivalent to the driving scale of the turbulence $l_0\simeq100\pc$ for spiral galaxies) 
with $512^3$ points. Exponentially growing magnetic fields are generated by such a flow via the fluctuation dynamo action.
The left-hand panel of Fig.~\ref{fig_particles} shows that the resulting magnetic field is spatially
intermittent being concentrated into filamentary structures.
A Gaussian random magnetic field with identical power spectrum is obtained from the intermittent one
by phase randomization (\cite{SSSBW17}). 
%------------------------------------------------------
%----------------------------------------------------
For an ensemble of particles in each of these fields, we then solve the 
particle equation of motion,
\begin{equation} \label{lf} 
\displaystyle\frac{\dd ^2\vec{r}}{\dd t^2}= \frac{v_0}{\Lar}\,\displaystyle\frac{\dd \vec{r}}{\dd t} \times \frac{\vec{b}}{\brms}\,, 
\end{equation}
where $\vec{r}$ is the particle's position at time $t$, $v_0$ is the particle's speed, $\Lar$ is the particle's 
Larmor radius (a parameter to control particle energy)
and $\vec{b}$ is the magnetic field assumed to be static at the time scales of interest. 
Initially, the particles are distributed uniformly in space with randomly directed velocities but fixed speed 
$v_0$. Once the diffusion sets in,
we calculate the coordinates of each particle modulo $2 \pi$, divide the numerical domain into $512^3$ cubes and count the number of cosmic ray particles in each cube to
obtain the cosmic ray number density as a function of time and position. Then we average over a long time to obtain the 
time-independent cosmic ray distribution $\ncr$. 
Since we neglect any energy losses, $\ncr$ represents a cosmic ray proton distribution.

The middle panel
of Fig.~\ref{fig_particles} shows 
$\ncr$ in an intermittent magnetic field. For both the Gaussian and intermittent fields, the magnetic field and cosmic rays are not correlated
and the correlation remains close to zero even when both distributions are averaged over any scale less than the box size. The probability density of $\ncr$  in both intermittent and Gaussian magnetic fields
is Gaussian at higher energies (particles with $\Lar > l_0$)
but develops a long, heavy tail at lower energies. The spatial distribution of 
low-energy cosmic rays ($\Lar \lesssim l_0$) is intermittent with numerous small-scale cosmic ray structures 
(middle panel of Fig.~\ref{fig_particles}). These structures are due to randomly distributed magnetic mirror traps. This is confirmed by 
the right-hand 
panel of Fig.~\ref{fig_particles} which shows a particle trajectory close to one such cosmic ray structures in the domain. The particle follows magnetic field lines 
(shown in dark grey) but is trapped between two magnetic mirrors. The number of such cosmic rays structures increases when a 
uniform magnetic field 
is included and decreases when 
pitch-angle scattering due to the unresolved magnetic fluctuations is added (Sect.~3.2 in \cite{SSWBS18}). 
However, for all cases
the cosmic rays and magnetic fields are not correlated at scales smaller than the box size. Furthermore, the magnetic field energy density ($b^2$) and cosmic ray ($\ncr$) distributions are statistically
independent of each other (Appendix C in \cite{SSWBS18}).
Assuming that this also applies to cosmic-ray electrons, this implies that Eq.~(\ref{syn}) can be written as
\begin{equation} \label{synbreak}
I = K \int_L \ncre b_{\perp}^2 \, \dd\ell = \frac{K}{L} \int_L \ncre \, \dd \ell  \int_L b_{\perp}^2 \, \dd\ell =  KL  \langle \ncre \rangle \langle b_{\perp}^2 \rangle\,.
\end{equation} 
Thus, one can use synchrotron intensity and the mean cosmic ray number density to obtain the average magnetic field strength.
%-------------------------------------------------------------------
\section{Cosmic rays as a fluid} \label{crfluid}
Cosmic rays exert pressure on the thermal gas which affects the magnetic field, which in turn controls the cosmic ray propagation. To include
this nonlinear effect, we solve the MHD equations
together with an advection-diffusion
equation for the cosmic ray fluid (see \cite{SBMS06} for equations, parameters and boundary conditions).
A random flow is driven at the box scale by an explicit force in the Navier--Stokes equation.
Cosmic rays are injected continuously at each time step but can be lost through the domain boundary. 
The magnetic field is first amplified exponentially, but then saturates due to the back-reaction of the Lorentz force on the flow. 
The mean cosmic ray energy also increases initially but then settles down to a steady value.
%------------------------------------------------------
\begin{figure}%[t]
       \centering
        \includegraphics[width=0.45\columnwidth,height=0.4\columnwidth]{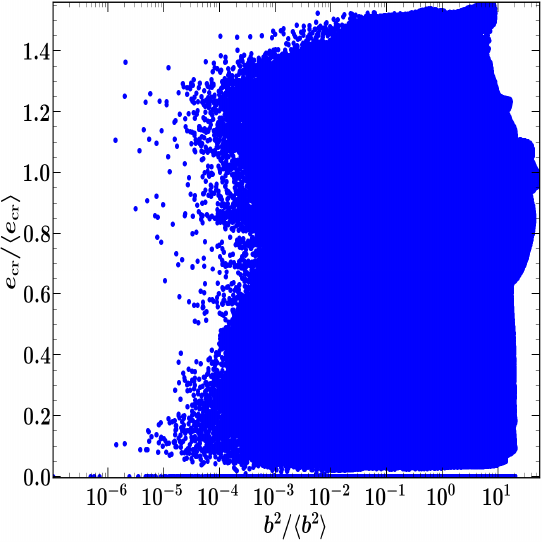} 
	\caption{The scatter plot for the normalized energy densities of cosmic rays $\ecr/\langle \ecr \rangle$ and magnetic fields  $b^2/\langle b^2 \rangle$ for the case
	with $\langle \ecr \rangle \approx \langle b^2 \rangle$ at the box scale. Even though both energy densities are equal when averaged over the 
	size of the domain, locally they are not correlated.}
        \label{fig_fluid}
\end{figure}%-------------------------------------------------------------
We vary the injection rate of cosmic rays such that, in the saturated stage, there
are three cases for the relation between the average cosmic ray $\ecr$ and 
magnetic field energy densities $b^2$: 
$\langle \ecr \rangle < \langle b^2 \rangle$, $\langle \ecr \rangle \approx \langle b^2 \rangle$ and $\langle \ecr \rangle > \langle b^2 \rangle$.
In all three cases, the cosmic ray energy density is not correlated with the magnetic field
energy density. Fig.~\ref{fig_fluid} confirms that the two quantities are uncorrelated even when $\langle \ecr \rangle \approx \langle b^2 \rangle$.
%--------------------------------------------------------------
\section{Conclusions} \label{conc}
Using both test-particle and fluid descriptions of cosmic rays, we have shown that the cosmic ray 
and magnetic field energy densities are not correlated on scales less than the driving scale of the turbulence ($l_0\simeq 100 \pc$ in spiral galaxies). 
Furthermore, the two quantities are statistically independent of each other, so the synchrotron intensity can be expressed as the product of the average cosmic ray number density
and average magnetic field strength. The presence of small-scale cosmic ray structures due to random magnetic traps can enhance the synchrotron
intensity locally. Such effects must be considered while analyzing high-resolution synchrotron observations of spiral galaxies. 
Our results do not exclude that energy equipartition may hold at scales larger than $l_0$.
%-----------------------------------------------

%\begin{discussion}
%\discuss{Massey}{}
%\discuss{van der Hucht}{}
%\end{discussion}

\end{document}